# Assessing the Impact of Retreat Mechanisms in a Simple Antarctic Ice Sheet Model Using Bayesian Calibration

Kelsey L. Ruckert[1], Gary Shaffer[2,3], David Pollard[1], Yawen Guan[4], Tony E. Wong[1], Chris E. Forest[1,5,6], Klaus Keller[1,6,7] *

**1** Earth and Environmental Systems Institute, The Pennsylvania State University, University Park, Pennsylvania, United States of America, **2** GAIA Antarctica, University of Magallanes, Punta Arenas, Chile, **3** Niels Bohr Institute, University of Copenhagen, Copenhagen, Denmark, **4** Department of Statistics, The Pennsylvania State University, University Park, Pennsylvania, United States of America, **5** Department of Meteorology, The Pennsylvania State University, University Park, Pennsylvania, United States of America, **6** Department of Geosciences, The Pennsylvania State University, University Park, Pennsylvania, United States of America, **7** Department of Engineering and Public Policy, Carnegie Mellon University, Pittsburgh, Pennsylvania, United States of America

* klaus@psu.edu

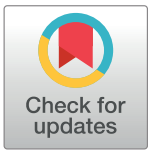

**Data Availability Statement:** All data underlying the study's findings are available from the Network for Sustainable Climate Risk Management Github Repository (https://github.com/scrim-network/Ruckertetal_DAIS_2016).

**Funding:** This work was partially supported by the National Science Foundation through the Network for Sustainable Climate Risk Management (SCRiM) under NSF cooperative agreement GEO-1240507 and the Penn State Center for Climate Risk Management. G. Shaffer was supported in part by Chilean ICM grant NC120066. Any opinions,

## Abstract

The response of the Antarctic ice sheet (AIS) to changing climate forcings is an important driver of sea-level changes. Anthropogenic climate change may drive a sizeable AIS tipping point response with subsequent increases in coastal flooding risks. Many studies analyzing flood risks use simple models to project the future responses of AIS and its sea-level contributions. These analyses have provided important new insights, but they are often silent on the effects of potentially important processes such as Marine Ice Sheet Instability (MISI) or Marine Ice Cliff Instability (MICI). These approximations can be well justified and result in more parsimonious and transparent model structures. This raises the question of how this approximation impacts hindcasts and projections. Here, we calibrate a previously published and relatively simple AIS model, which neglects the effects of MICI and regional characteristics, using a combination of observational constraints and a Bayesian inversion method. Specifically, we approximate the effects of missing MICI by comparing our results to those from expert assessments with more realistic models and quantify the bias during the last interglacial when MICI may have been triggered. Our results suggest that the model can approximate the process of MISI and reproduce the projected median melt from some previous expert assessments in the year 2100. Yet, our mean hindcast is roughly 3/4 of the observed data during the last interglacial period and our mean projection is roughly 1/6 and 1/10 of the mean from a model accounting for MICI in the year 2100. These results suggest that missing MICI and/or regional characteristics can lead to a low-bias during warming period AIS melting and hence a potential low-bias in projected sea levels and flood risks.

findings, and conclusions or recommendations expressed in this material are those of the authors and do not necessarily reflect the views of the National Science Foundation or other funding entities. The funders had no role in study design, data collection and analysis, decision to publish, or preparation of the manuscript.

**Competing Interests:** The authors have declared that no competing interests exist.

# Introduction

Coastal areas are at risk to sea-level rise, and will be more so if the marine part of the Antarctic ice sheet (AIS) were to collapse. A disintegration of the marine part of the AIS would raise global mean sea level by more than three meters and hence drive an increase in flooding vulnerability in many coastal areas [1]. Basic physics, the paleo-record, as well as model simulations suggest that such a disintegration could follow a highly nonlinear, relatively abrupt, and hysteresis type threshold response [2–5]. Previous studies suggest that anthropogenic greenhouse gas emissions could trigger a disintegration of the marine part of the AIS [2, 6–8], and implies that assessing these potential increased risks is important.

Hindcasts and projections of AIS dynamics are deeply uncertain. This uncertainty stems in part from the complexity of the coupled AIS / ocean / atmosphere systems combined with relatively sparse data and the severely limited ability to resolve relevant processes in current models [3, 5, 9, 10]. For example, consider the highly vulnerable marine part of the AIS. The marine part of the AIS is vulnerable to anthropogenic climate change because it is grounded below the sea level and in direct contact with the ocean via ice shelves [2, 8, 11]. Ocean-ice interactions can in part drive AIS dynamics, for example, through melting of ice shelves and calving cliffs [5, 12]. For the part of the ice sheet grounded below sea level, this melting and calving can lead to a runaway retreat due to threshold behavior, known as Marine Ice Sheet Instability (MISI) [2, 3, 5]. MISI is a runaway retreat of the grounding line as warming temperatures increase both the water depth and ice flux at the grounding line [8]. Once started, this mechanism will continue (even in the absence of external forcings) until the grounding line reaches an upward-sloping bed or enough buttressing is exerted to stabilize the grounding line on a retrograde slope [13]. Previous studies suggest that the additional process of Marine Ice Cliff Instability (MICI) may facilitate MISIs and therefore ice sheet disintegration [2, 3, 5, 8, 12]. MICI is defined as a weakening or structural failure of shear ice cliffs as warming temperatures increase crevasses and reduce the maximum supported cliff heights [8].

Here, we calibrate a previously published and relatively simple model of AIS volume loss [14] to reproduce a hindcast period from the last interglacial to the present using a pre-calibration method and a Bayesian inversion. The Bayesian inversion accounts for the heteroskedastic nature of the data and the model accounts for MISI (Fig 1). However, the model neglects the effects of MICI and is unable to capture regional characteristics of the ice sheet. To approximate the effects of missing MICI, we compare our projections to physically more realistic models [8, 15–18] and quantify bias during the last interglacial period, when MICI was potentially triggered. In the results section, we show that our projections are comparable to some previous expert assessments [4, 15–18], yet unable to account for the sizable melt generated from a model considering MICI [8] and that our mean hindcast is unable to reproduce the estimated mean during the last interglacial. In our analysis, we deduce that missing MICI is one possible mechanism that can lead to underestimation of warming period (i.e., the last interglacial and the year 2100) melting in the AIS model.

# Methods and data

## Antarctic ice sheet model

We adopt the DAIS (Danish Center for Earth System Science Antarctic Ice Sheet) model as a starting point for our analysis [14]. This model has been previously described in detail (see for example, [14]), hence we provide just a brief overview. The DAIS model considers three important changes in the ice sheet: ice sheet volume, volume loss in sea-level equivalence (SLE), and ice sheet radius. In this model, the volume for the entire AIS changes under four

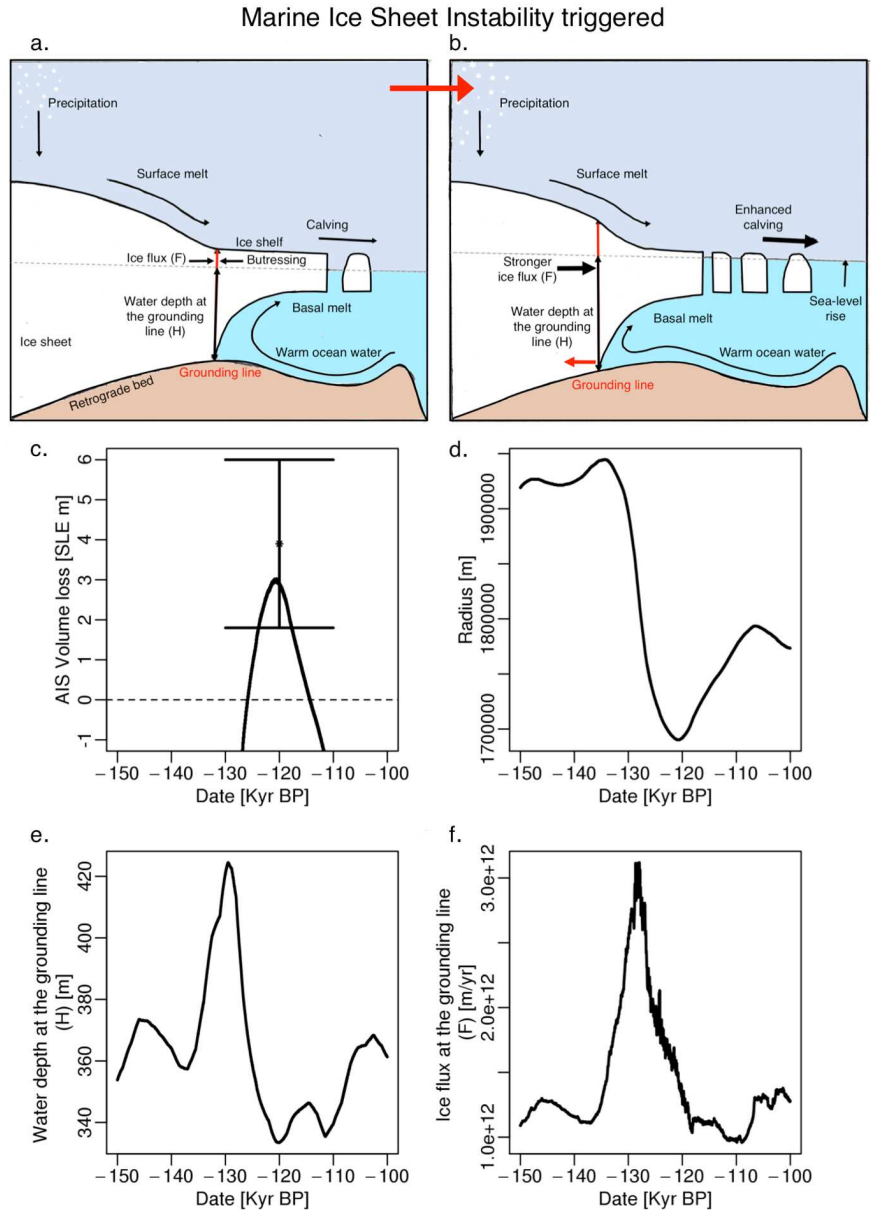

**Fig 1. Indication of Marine Ice Sheet Instability (MISI) triggered in the model during the last interglacial (130 to 110 kyr BP).** The illustration in panel a and b depict a schematic cross section of an ice sheet approaching the MISI. Ice flows from the grounded ice to the floating ice shelf. As surface runoff, calving, and basal melt increase (a to b) the flow across the grounding line increases and the grounding line retreats rapidly further inland unabated (red arrow). In the model, MISI occurs during the last interglacial period when there is (c) deglacial retreat as the (d) radius decreases, the (e) water depth at the grounding line increases, and the (f) ice flux across the grounding line strongly increases.

doi:10.1371/journal.pone.0170052.g001

mechanisms: (i) ice sheet growth from precipitation, (ii) accumulation from precipitation minus melt runoff, (iii) accumulation from precipitation minus basal melt, and (iv) ice sheet disintegration from precipitation minus melt runoff and basal melt. The simplicity of the model stems from neglecting key processes and threshold responses. For example, the model misses the link of warming temperatures to hydro-fracturing of buttressing ice shelves, collapse of ice cliffs, and abrupt changes in melting rates by surpassing tipping points.

Additionally, the simplicity of the model treats the AIS as a whole rather than capturing the regional characteristics of glaciers and subglacial basins, which would help trigger the MISI (see the section "Potential explanation for the discrepancy" for more details). However, the model quantifies the ice flux (the rate at which the ice flows towards the sea) at the grounding line (location where the ice shelf begins to float; averaged for the entire AIS), the water depth at the grounding line (distance between the sea level and the grounding line; averaged for the entire AIS), and the relationship between ice sheet volume and ice sheet radius [14] (Fig 1). In the model, ice sheet volume relates to the radius based on an undisturbed bed profile referenced to present day volume and radius taking into account isostatic adjustment and its effect on the displacement of seawater by ice [11, 14]. Despite the drawbacks, the simplicity and computational efficiency of the DAIS model enables us to implement a Bayesian data-model fusion and to quantify uncertainties with a focus on the tails of the distributions.

### Data

**Forcings.** The ice sheet forcings in the hindcast period span from 240 kyr BP to the year 1997 [14]. These forcings include Antarctic temperature reduced to sea-level, global mean sea level, global mean sea-level rate, and high latitude subsurface ocean temperature.

For projections beyond 1997, we generate forcings with the CNRM CM5 RCP8.5 temperature scenario (as obtained from the CMIP5 model output archive, http://cmip-pcmdi.llnl.gov/cmip5/). This model simulation was chosen because both surface air and ocean temperatures were projected to the year 2300 using the extended RCP8.5 scenario. The future forcings are created using details and instructions specified in a previous analysis [14]. The Antarctic temperature anomaly is adjusted to a present-day (1961–1990) mean temperature of -18°C and projected over a 1°lat./long. Antarctica land mask [14]. We generate future global mean sea level and sea-level rates with an empirical global mean sea-level model [19] calibrated using historical NASA GISS global mean temperatures [20, 21] and global mean sea-level observations [14] by minimizing the sum of the absolute residuals using a global optimization method [22]. The high latitude subsurface ocean temperature is the volume averaged temperature from 200–800 m depth between 52–70°S [14]. In addition to the business-as-usual temperature scenario, the ocean temperatures are projected with the Antarctic temperature forcing and adjusted to a present-day mean temperature of 0.72°C.

**Observational constraints.** We use four observational constraints of volume loss in SLE to fit the model: (i) the last interglacial (LIG; ∼120 kyr BP), (ii) the last glacial maximum (LGM; ∼20 kyr BP), (iii) the mid-Holocene (MH; ∼6 kyr BP), and the instrumental period (1992–2011) (S1 Table). Each observational constraint is based on previously published work and interpreted as a 95% confidence interval. These ranges include: 1.8–6.0 m (LIG), -6.9–-15.8 m (LGM), -1.25–-4.0 m (MH), and 1–2.9 mm (in the year 2002). Note that the instrumental period constraint is relative to the year 1992, whereas the other constraints are relative to the 1961–1990 period. A detailed discussion of the constraints is provided in the Supporting Information (S1 Text and S1 Table).

### Calibration methods

We use two separate calibration stages: (i) a pre-calibration (e.g., [23]) using Latin hypercube sampling (LHS) and (ii) a full Bayesian inversion based on a Markov chain Monte Carlo (MCMC) algorithm. We compare these two methods to assess the potential improvement in the hindcasts and parameter estimates associated with moving from the simpler pre-calibration to a full Bayesian inversion and to provide a check for the Bayesian inversion.

Table 1. Comparison of the percentage of runs passing through individual constraints, all the constraints, or no constraints for the Pre-calibration (n = 1.3 × 10³) versus full Bayesian inversion method (subset n = 3.5 × 10³).

| Method | No constraints (%) | Last interglacial (%) | Last glacial maximum (%) | Mid-Holocene (%) | Instrumental period (%) | All constraints (%) |
|---|---|---|---|---|---|---|
| Pre-calibration | 100 | 53 | 30 | 50 | 5 | 1 |
| Bayesian inversion | 100 | 74 | 76 | 72 | 8 | 3 |

doi:10.1371/journal.pone.0170052.t001

**Pre-calibration.** We first implement pre-calibration using a LHS technique [24, 25]. We generate $1.3 \times 10^3$ samples from the 13-dimensional model parameter space. We use uniform prior distributions for each physical model parameter and an inverse gamma distribution for the statistical parameter $\sigma^2_P$ [26, 27]. The parameters $\sigma^2_P$ and $\sigma^2_I$ approximate the effects of observational errors, unresolved internal variability, and model errors for the paleo and instrumental period (discussed below). LHS divides each parameter distribution into $1.3 \times 10^3$ equally probable intervals and then draws $1.3 \times 10^3$ sample points. We employ a maxi-min LHS, which optimizes the sample by maximizing the minimum distance between parameter values [25, 28]. Using this method, we generate $1.3 \times 10^3$ parameter combinations and determine which parameter sets satisfy individual observational constraints (Table 1; hindcasts and projections shown in S1 and S2 Figs).

**Markov chain Monte Carlo (MCMC).** We implement a Bayesian inversion technique using a MCMC algorithm [29–31]. We represent the observational data of AIS SLE by $y_t$, the model output as $f(\boldsymbol{\theta}, t)$, the unknown model parameters $\boldsymbol{\theta}$, an unknown residual term $R_t$, and time $t$ over a hindcast period of 240 kyr BP to 1997. We approximate the AIS volume loss observations as the sum of the model output plus an unknown residual term,

$$y_t = f(\boldsymbol{\theta}, t) + R_t. \qquad (1)$$

The observations, model output, and residuals are vectors from 1 to N (N = 4 in this case, since there are four observational constraints) such that $\boldsymbol{y} = (y_1, \ldots, y_N)^t$, $\boldsymbol{f}(\boldsymbol{\theta}, t) = (f(\boldsymbol{\theta}, t_1), \ldots, f(\boldsymbol{\theta}, t_N))^t$, and $\boldsymbol{R} = (\delta_1 + \epsilon_1, \ldots, \delta_N + \epsilon_N)^t$. Because thousands of years separate the observational constraints, we neglect the temporal correlation. Hence, we model the residuals $R_t$ as drawn from an independent and identically distributed normal distribution $\delta_t$ with an added observation error $\epsilon_t$,

$$R_t = \delta_t + \epsilon_t. \qquad (2)$$

This is a simple model for the data model discrepancy since accounting for independent and identically distributed residuals ignores the potentially complex autocorrelation structure [32] of AIS trends. The autocorrelation structure can be modeled from year to year if more data are available [32]. In our case, our simple model seems appropriate, because there are four data points with large time spans between observations. We hence use models of white noise, $\delta_t$ that are independent normally distributed with different variance for the paleo and instrumental record, for instance, $\delta_t \sim N(0, \sigma^2_P)$ for t = 1, 2, 3 and $\delta_t \sim N(0, \sigma^2_I)$ for t = 4. The observation error $\epsilon_t$ is heteroskedastic consisting of a normal distribution with zero means and a time dependent variance, $\epsilon_t \sim N(0, \tau^2_t)$. The white noise $\boldsymbol{\delta} = (\delta_1, \ldots, \delta_N)^t$ approximates the model structural uncertainty and internal climate variability whereas the time dependent variance of $\boldsymbol{\epsilon} = (\epsilon_1, \ldots, \epsilon_N)^t$ is implemented by substituting in the known measurement error of each observational constraint. This implementation is important because ignoring the potentially complex error structure can result in overconfident parameter estimates, hindcasts, and projections (e.g., [32]).

The data assimilation technique follows Bayes' theorem [33], where the posterior probability $\pi$ of observing the parameters $\Theta$ given the data $y$ is proportional to the likelihood $L$ of the data given the parameters times the prior probability distribution of the parameters,

$$\pi(\Theta \mid y) \propto L(y \mid \Theta) \times \pi(\Theta). \quad (3)$$

After specifying the prior parameter distributions (S3 Fig) and the likelihood function, the MCMC algorithm samples from the posterior distribution [29–31, 34]. We construct the likelihood function, assuming that the observational data $y$ consists of the model output $f(\theta, t)$ plus an unknown residual term $R$ (Eqs 1 and 2). The uncertain parameters $\Theta$ include both the unknown model parameters $\theta$ and the unknown statistical parameters $\sigma^2_P$ and $\sigma^2_I$ (S3 and S4 Figs). To derive the likelihood of the data given the parameters $L(y|\Theta)$, we find the residuals $R \sim N(0, \Sigma)$;

$$\Sigma = Var(R) = \begin{pmatrix} \sigma^2_P + \tau^2_1 & 0 & 0 & 0 \\ 0 & \sigma^2_P + \tau^2_2 & 0 & 0 \\ 0 & 0 & \sigma^2_P + \tau^2_3 & 0 \\ 0 & 0 & 0 & \sigma^2_I + \tau^2_4 \end{pmatrix}. \quad (4)$$

The likelihood functions is then:

$$L(y \mid \Theta) = \left(\frac{1}{\sqrt{2\pi}}\right)^N \times |\Sigma|^{\frac{-1}{2}} \exp\left\{\frac{-1}{2}[y - f(\theta, t)]^\tau \Sigma^{-1}[y - f(\theta, t)]\right\}. \quad (5)$$

We adopt uniform priors for the model parameters and an inverse gamma prior for the statistical parameter $\sigma^2_P$ ($\alpha = 2$; $\beta = 1$) (S3 Fig). The inverse gamma prior for the paleo period $\sigma^2_P$ is a vague prior with a heavy tail. Due to convergence difficulties, an inverse gamma prior could not be used for $\sigma^2_I$; a uniform prior of 0–0.0004 was used instead (S3 Fig). The adaptive MCMC uses Metropolis-Hastings updates with joint normal proposal [29, 30, 34]. Following standard practice, we use a large number of samples from the MCMC chain with $8 \times 10^5$ iterations, minimize effects of the initial parameter guess in the MCMC samples by removing a 4% burn-in, and thin the chains for analysis [31]. Visual inspection and the potential scale reduction factor suggest the Markov chains are well mixed and converged [35]. We drive the calibrated model with future forcings to project smooth simulations $f(\theta, t)$ of AIS volume loss in SLE. Following Eq (1), the smooth simulations are superimposed with the process noise to account for the internal variability and heteroskedasticity not captured by the model. We superimpose the smooth simulations with the paleo process noise from 240 to 5 kyr BP and superimpose the instrumental process noise from 1961 to the year 2300. We apply a linear regression from the paleo to the instrumental process noise to simulate a smooth transition period from 5 kyr BP to the year 1960. The simulations with process noise represent a probabilistic distribution of AIS volume loss in SLE and the Markov chains represent the empirical estimate of the joint parameter posterior (S3 and S4 Figs).

## Results

### A full Bayesian inversion improves hindcasts relative to pre-calibration

Moving from the pre-calibration method to a full Bayesian inversion improves the fit of the model to the observational constraints. As perhaps expected, the more sophisticated Bayesian calibration method increases the percentage of samples passing through each individual observational constraint (Table 1). More importantly, we assess the hindcast skill of the model fit with the root-mean-square error (RMSE) between the mean and median model fit and the observational constraints. The Bayesian inversion improves the hindcast skill by reducing the root-mean square error from roughly 5.5 and 3.8 m (mean and median pre-calibration RMSE) to 0.8 m (Bayesian inversion RMSE for both mean and median).

### The calibrated model captures MISI during the last interglacial period

The mean hindcast produces enough melt, decrease in radius, and increase in ice flux to suggest the model triggers MISI during the LIG period. MISI occurs when oceanic and atmospheric warming causes the ice flux at the grounding line to increase and the grounding line to retreat unabated until the grounding line reaches an upsloping bed or temperatures cool enough to reform and stabilize the buttressing ice shelf (Fig 1a and 1b) [8, 13]. During the LIG, the AIS melts by roughly 2.9 m in SLE (mean) causing a decrease in the ice sheet radius (Fig 1c and 1d). As this occurs, the water depth and the ice flux at the grounding line increase suggesting retreat of the grounding line (Fig 1e and 1f). The combination of the increased melt, ice flux, and water depth at the grounding line not only demonstrates that the model captures MISI, but that the model can produce enough melt to simulate a loss of the marine ice sheet (i.e., a contribution to global mean sea-level rise of more than three meters; [1]).

### The calibrated model partially passes the hindcast test, with evidence for structural discrepancies

The hindcast test reveals some ability to reproduce observational constraints as well as structural model discrepancies. The model does reasonably well at reproducing the width of observational constraint (95% confidence interval) during the LGM, MH, and the instrumental period (Figs 2 and 3). However, our model fit does not reproduce the mean and the top percentage of melting during the LIG. During the LIG, our calibrated model hindcasts roughly 2.9 m, or between -0.1 and 5.9 m (mean and 95% credible interval) of AIS contribution to sea level. These results indicate that our mean hindcast is roughly 26% (1 m) lower than the LIG observed data (Figs 2 and 3a).

### Future sea-level contributions diverge from projections considering MICI

Our projection, in the year 2100, produces results that are comparable to previous expert assessments [15–18]. These previous expert assessments consider the sensitivity or vulnerability of glacier collapse (e.g. Pine Island and Thwaites) within the next century, project changes in surface mass balance and potential rapid ice sheet dynamics, or use a sub-grid interpolation of basal melting at the grounding lines [15–18]. Our calibrated model (in 2100) has an AIS contribution of roughly 0–0.3 m (90% credible interval). In comparison, the previous expert assessments [15–18] generally produce a similar or narrower uncertainty range for the 90% confidence/credible interval or low to high estimate (Fig 3e and Table 2). Overall, our median projection is comparable, differing by roughly 3–5 cm (Fig 3e and Table 2).

Additionally, we compare our projections in the year 2100 to some non-model expert assessments [4, 9, 36]. In contrast to the comparisons with the model expert assessments [15–18], the non-model expert assessments produce a wider [9, 36] or a narrower [4] uncertainty range for the 90% confidence/credible interval. The differences between our results versus the non-model expert assessments are likely due to several reasons. For instance, the non-model expert assessments produce their estimate based on an interpretation of various model studies and scaled to account for dynamical effects, based on formal expert elicitation by conducting expert interviews, and by using maximal kinematic constraints [4, 9, 36]. None of the non-

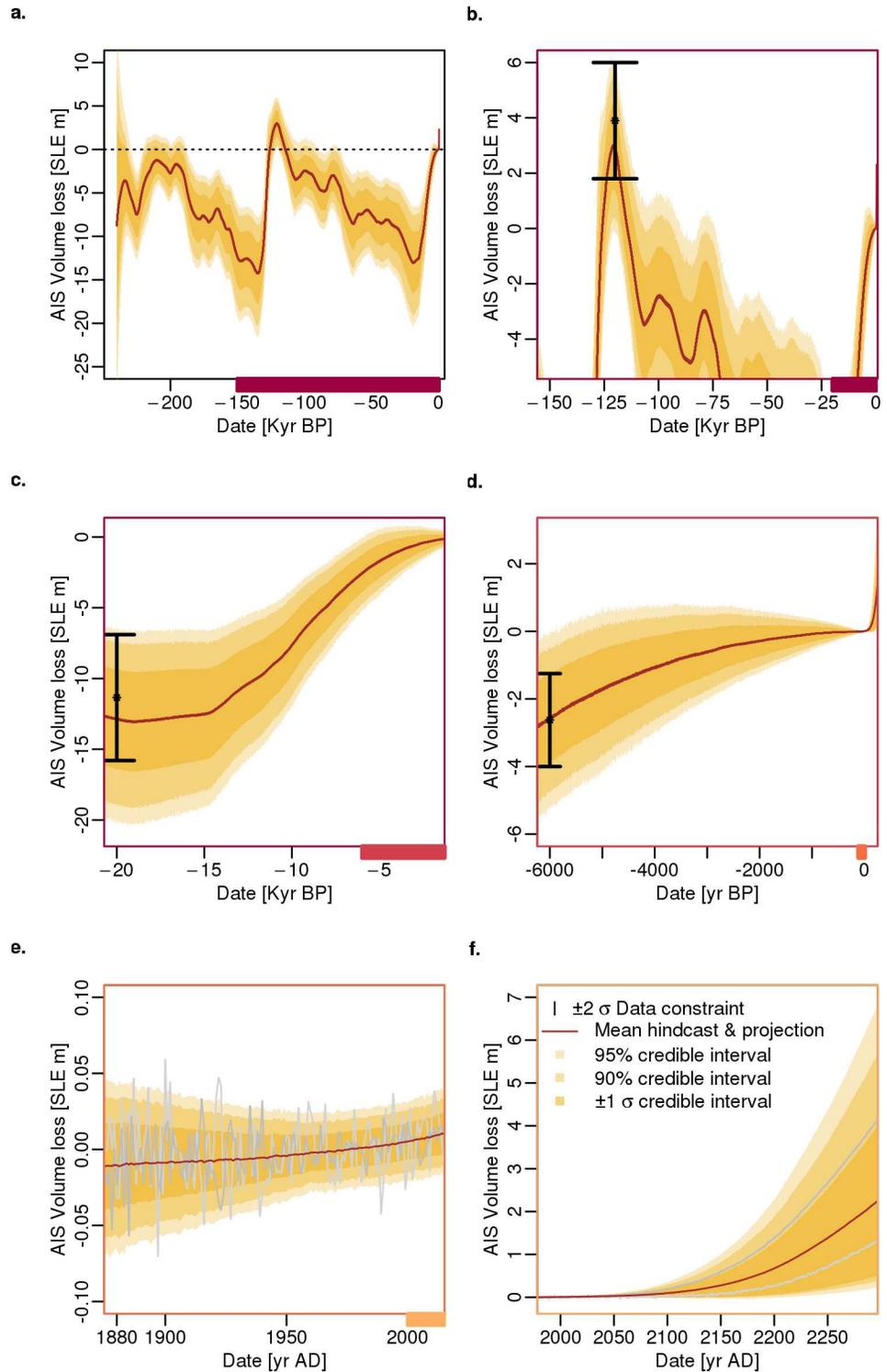

**Fig 2. Hindcasts and projections of global mean equivalent sea-level (SLE) rise from the Antarctic ice sheet (AIS).** The (a) full time series is broken down to focus on the (b) last interglacial, (c) last glacial maximum, (d) mid-Holocene, (e) the instrumental period, and (f) projections to the year 2300. Shown are the ± 1σ, 90, and 95% credible intervals (tan to gold), the fitted mean (brown), observational constraints (bars), and two random runs in grey.

doi:10.1371/journal.pone.0170052.g002

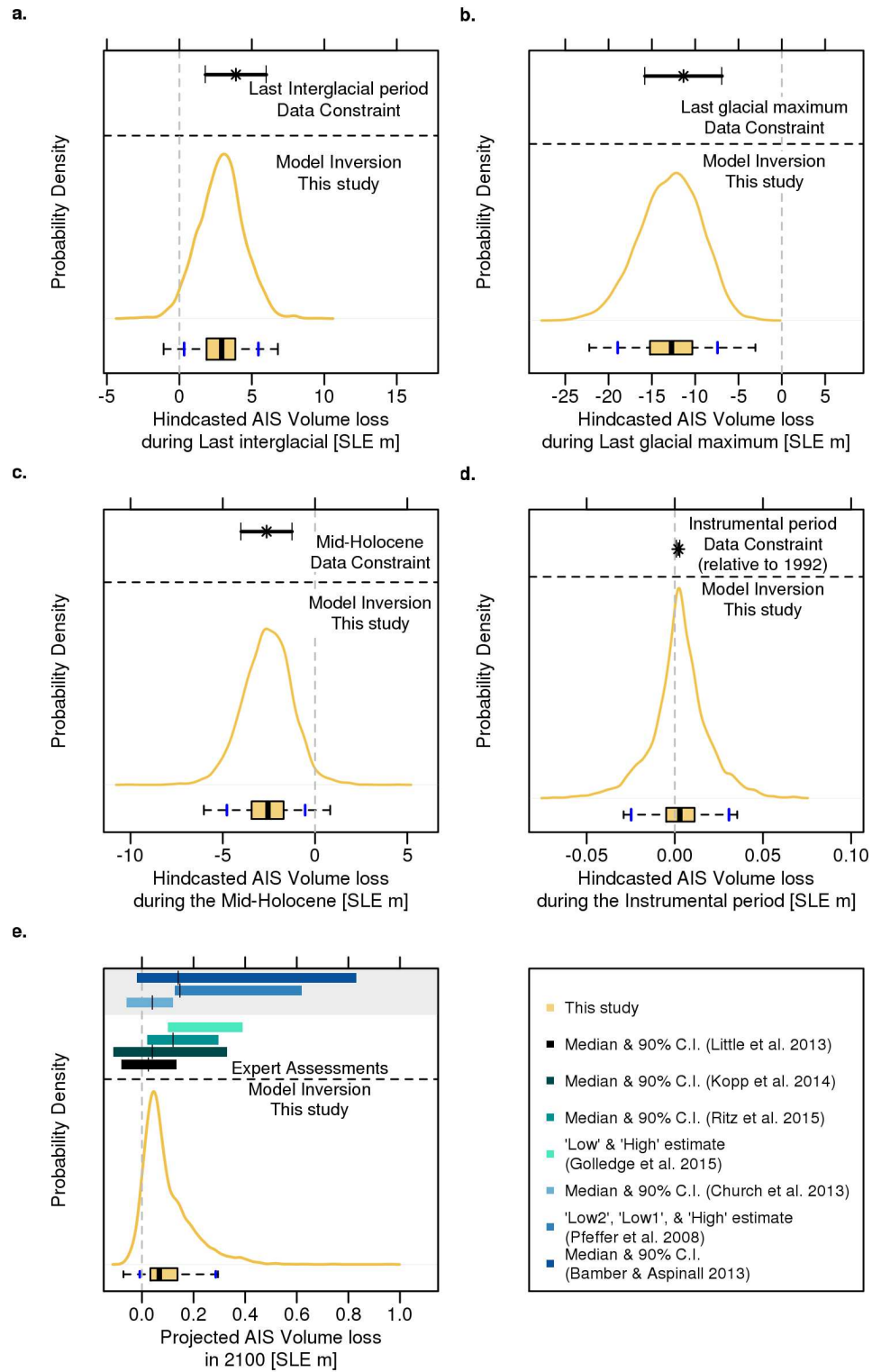

**Fig 3. Comparison of the fitted Antarctic ice sheet (AIS) volume loss presented as probability density functions versus data constraints and expert assessments.** The gold line is the AIS volume loss hindcast/projection in this study. The black range represents the reconstructed AIS observational constraints and the different colored horizontal bars are the 90% confidence/credible intervals with their estimated medians (|) from expert assessments. The bars in shades of green are from model assessments projected with the RCP8.5 or similar scenario. Bars in shades of blue, highlighted in grey, are from non-model assessments.

doi:10.1371/journal.pone.0170052.g003

Table 2. Comparison of projections in the year 2100 to previous expert assessments.

| Studies | Mean (m) | Median (m) | Low-high (m) | 90% CI (m) | ± 1σ (m) |
| --- | --- | --- | --- | --- | --- |
| This study | 0.09 | 0.07 | - | -0.01–0.29 | 0–0.19 |
| Little et al. 2013 | - | 0.024 | - | -0.08–0.133 | - |
| Kopp et al. 2014 | - | 0.04 | - | -0.11–0.33 | - |
| Golledge et al. 2015 | - | - | 0.1–0.39 | - | - |
| Ritz et al. 2015 | - | 0.119 | - | -0.02–0.296 | - |
| DeConto and Pollard 2016 (5–15m Pliocene target) | 0.64 | - | - | - | 0.15–1.13 |
| DeConto and Pollard 2016 (10–20m Pliocene target) | 1.05 | - | - | - | 0.75–1.35 |
| Pfeffer et al. 2008[a] | - | - | 0.128–0.146–0.619[b] | - | - |
| Bamber and Aspinall 2013[a] | - | 0.14 | - | -0.02–0.83 | - |
| Church et al. 2013[a] | - | 0.04 | - | -0.06–0.12 | - |

[a] Non-model assessments.
[b] Displays the low2, low1, and high estimate.

doi:10.1371/journal.pone.0170052.t002

model expert assessments are calibrated to a RCP8.5 or similar scenario, making comparison to these assessments difficult.

Although our model calibration results in projections that are comparable to previous expert assessments based on models [15–18], our projection is unable to account for the sizable melt generated from a physically more realistic model considering hydro-fracturing and MICI [8] (Fig 4). Our calibrated model without MICI projects an AIS contribution of roughly 0.09 m (mean) and 0–0.2 (± 1σ) in the year 2100 (Fig 4 and Table 2). In comparison, our results, which neglect MICI, miss 96 and 100% of the ± 1σ projections produced in the physically more realistic model [8] (Fig 4). Furthermore, our mean projection, in 2100, is roughly 1/6 and 1/10 (0.55 and 0.96 m smaller) of the mean produced in the physically more realistic model [8] (Fig 4 and Table 2).

## Potential explanations for the discrepancy

The missing key processes and characteristics in the model such as the effect of MICI, the capturing of regional characteristics, and effect of hydro-fracturing of buttressing ice-shelves are potential explanations for this discrepancy. The calibrated model produces discrepancies during periods warmer than today (i.e., LIG and 2100). During the LIG, global mean temperatures were roughly 2°C above today and evidence indicates a potential collapse of the marine ice sheet [1, 37, 38]. This rise in temperature is consistent with projections over the next few centuries [37, 38]. As discussed in the introduction, melting and calving from warming temperatures can trigger rapid runaway retreat of the ice sheet grounded below sea level, MISI (Fig 1a and 1b). MISI is highly dependent on the regional characteristics of subglacial basins. For instance, MISI can occur for the West Antarctic ice sheet based on the stability of Pine Island Glacier, Thwaites Glacier, and other glaciers. Yet, high volume losses will be defined by thresholds for the marine part of the East Antarctic ice sheet. A potential mechanism to trigger such losses faster is the MICI mechanism, which can help facilitate MISI in such marine basins and therefore ice sheet disintegration. Despite the models ability to capture MISI, the model neglects the effects of MICI and the ability to capture regional characteristics of the ice sheet. Hence missing MICI as well as the inability to estimate regional characteristics are two potential causes of the reduced melting.

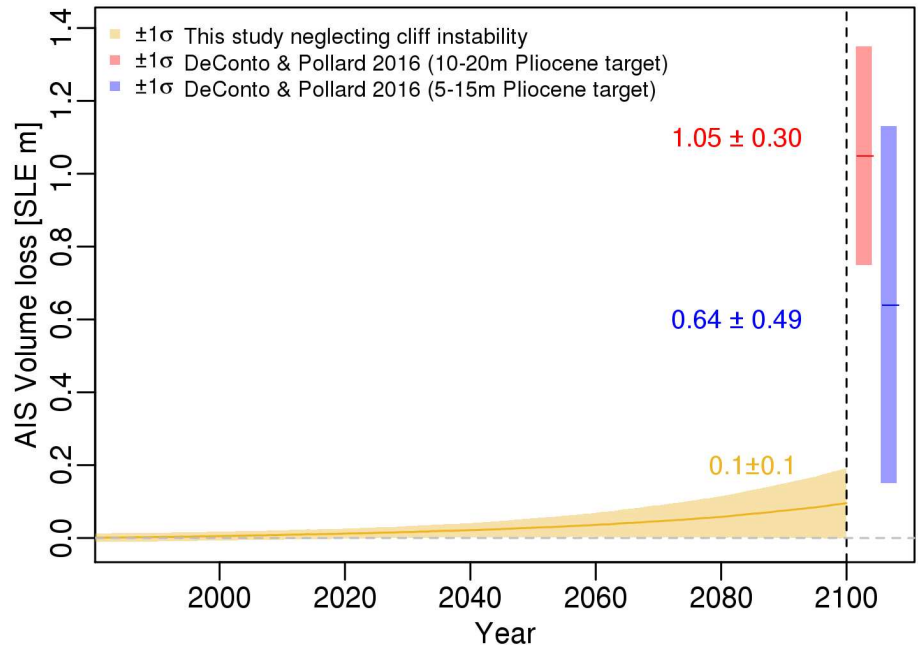

**Fig 4. Comparison of the fitted Antarctic ice sheet volume loss to projections considering Marine Ice Cliff Instability (MICI).** The gold line and envelope represent the mean and ± 1σ credible interval produced in this study neglecting cliff instability. The red and blue line and bar represent the mean and the ± 1σ projection in the year 2100 from a physically more realistic model [8], which considers key processes in the model. The red and blue projections differ by choice of observational constraint during the Pliocene.

doi:10.1371/journal.pone.0170052.g004

## Discussion and caveats

This analysis adopts a relatively simple model structure and neglects several potentially important sources of uncertainty. The modeling choice is motivated by the goal to produce a transparent analysis, but this simplicity also points to caveats and future research needs. For example, we analyze the question whether and how much neglecting MICI can reduce AIS melt during warming periods by comparing our results to projections from a physically more realistic model [8], a model accounting for MICI. A more refined (but also much more involved) approach to quantify this bias would be to calibrate a model with and without MICI. Additionally, we make expert judgments of the prior range for each parameter given that there are only four data points, but 13 parameters. Imposing hard bounds potentially cuts off the tails of the posterior parameter densities, however, providing the model with far too much flexibility given the small number of data points could lead to physically unrealistic results. Furthermore, our analysis neglects the impacts of many other uncertainties (e.g., about the paleo and future forcings). For example, the sea-level curve preceding the LIG might be considerably revised/improved using more recent work (see for example, [39, 40]). Despite these caveats, the relative simplicity of the analysis enables the approximation of the effect of missing MICI on AIS melt in the model.

## Conclusion

We calibrate a simple AIS model (that does not include a cliff instability mechanism nor is able to capture regional characteristics) with observational constraints over the past 240,000 years using a Bayesian inversion considering the heteroskedastic nature of the data. Using the hindcasts and projections, we compare our results to those from a pre-calibration method and expert assessments with potentially more realistic models. We approximate how neglecting fast processes (i.e., the MICI mechanism) in an AIS model can lead to biases in the AIS hindcasts and projections during warming periods. For the specific example considered, we show how missing MICI produces a lower mean hindcast (roughly 26% or 1 m smaller) during the LIG, a period when the marine ice sheet is suggested to have deglaciated. Additionally, the model is unable to account for roughly 96 and 100% of future AIS contributions predicted by a physically more realistic model accounting for MISI, MICI, and hydro-fracturing yet reproduces the projected median melt in other expert assessments in the year 2100. Overall, accounting for retreat mechanisms can potentially increase warming period AIS melt and reduce model discrepancy.

## Supporting Information

**S1 Text. Discussion of constraints used for calibrating the DAIS model.**
(PDF)

**S1 Table. Range, median, and error estimates for the four constraints used to calibrate the DAIS model.** A traceable account of previous studies used to create the constraints is listed in the references.
(PDF)

**S2 Table. Information about the CMIP modeling group whose model output was used in this study archived at http://cmip-pcmdi.llnl.gov/cmip5/.** This table was accessed and modified on 13 June 2016 from http://cmip-pcmdi.llnl.gov/cmip5/docs/CMIP5_modeling_groups.docx.
(PDF)

**S1 Fig. Comparison of hindcasts and projections of Antarctic ice sheet volume loss pre-calibrated with different constraints.** Shown are the realizations pre-calibrated with the Last interglacial constraint (green), Last glacial maximum constraint (turquoise), mid Holocene constraint (yellow), instrumental period constraint (blue), all the constraints (navy), and no constraints (gray). The brown line represented the optimized fit and the bars are the observational constraints.
(TIFF)

**S2 Fig. Comparison of the pre-calibrated Antarctic ice sheet (AIS) volume loss hindcast/projection during (a) the last interglacial and (b) the year presented as probability density functions versus data constraints and expert assessments.** The different coloured probability density functions represent realizations using different constraints (last interglacial (green), last glacial maximum (turquoise), mid-Holocene (yellow), instrumental period (blue), all the constraints (navy), and no constraints (gray)). The black range represents the reconstructed AIS contribution to global sea level during the last interglacial period. The different coloured lines are the 90% confidence/credible intervals with their estimated medians (|) from previous AIS studies. The bars in shades of green are from model assessments projected with the RCP 8.5 or similar scenario. Bars in shades of blue, highlighted in grey, are from non-model assessments.
(PNG)

**S3 Fig. Marginal probability density functions of the estimated parameters.** The horizontal axis range represents the lower and upper bounds of the uniform prior distributions. The variance parameter $\sigma^2_P$ uses an inverse gamma prior distribution ($\alpha = 2$, $\beta = 1$) with a minimum

PLOS ONE | Assessing the Impact of Antarctic Retreat Mechanismsvalue of 0 and an infinite upper bound. The table indicates the estimated parameters their prior ranges.
(PDF)

**S4 Fig. Relationship of Bayesian inversion estimated model parameters shown as a pairs plot.** Values increase in likelihood from blue to tan.
(PNG)

## Acknowledgments

We thank A. Bakker, G. Garner, and R. Fuller for assistance on speeding up the DAIS model and for testing the reproducibility of the code. Additionally, we thank P. Oddo and P. Applegate for Latin hypercube sampling assistance. Also, we thank Varada Vaidya for producing future forcings. We thank M. Haran, D. Diaz, and B. Forest for valuable inputs. This work was partially supported by the National Science Foundation through the Network for Sustainable Climate Risk Management (SCRiM) under NSF cooperative agreement GEO-1240507 and the Penn State Center for Climate Risk Management. G. Shaffer was supported in part by Chilean ICM grant NC120066. Any opinions, findings, and conclusions or recommendations expressed in this material are those of the authors and do not necessarily reflect the views of the National Science Foundation or other funding entities. We also acknowledge the World Climate Research Programme's Working Group on Coupled Modelling and thank the climate modeling groups that participated in the Coupled Model Intercomparison Project Phase 5 (CMIP5; http://cmip-pcmdi.llnl.gov/cmip5/), which supplied the climate model output used in this paper (listed in S2 Table). The US Department of Energy's Program for Climate Model Diagnosis and Intercomparison in partnership with the Global Organization for Earth System Science Portals, provides coordinating support for CMIP5. Data and codes are available through the corresponding author and will be available via Github at https://github.com/scrim-network/Ruckertetal_DAIS_2016 upon publication.
## Author Contributions

**Conceptualization:** KLR GS DP YG TEW CEF KK.

**Data curation:** KLR.

**Formal analysis:** KLR.

**Funding acquisition:** KK GS.

**Investigation:** KLR.

**Methodology:** KLR GS DP YG TEW.

**Project administration:** KK.

**Resources:** KK.

**Software:** KLR.

**Supervision:** KK.

**Validation:** TEW.

**Visualization:** KLR.

PLOS ONE | DOI:10.1371/journal.pone.0170052   January 12, 2017                                                    13 / 15

**Writing – original draft:** KLR.

**Writing – review & editing:** KLR GS DP YG TEW CEF KK.